%% file: main.tex
\documentclass{ppig}
\usepackage{epsfig}
\usepackage{booktabs}
\usepackage{ucs}
\usepackage[utf8x]{inputenc}

\usepackage{natbib}
\usepackage{xcolor}
\usepackage{url}
\usepackage[colorlinks=true,citecolor=black]{hyperref}
\newcommand{\el}[1]{[...]}

\renewenvironment{quote}
           {\vspace{-0.3cm}
           \list{}{\rightmargin\leftmargin}%
            \item\relax\small\itshape\color{darkgray}\ignorespaces}
           {\unskip\unskip\endlist
           \vspace{-0.3cm}}

\title{What is it like to program with artificial intelligence?}

\author{Advait Sarkar \\
  Microsoft Research \\
  University of Cambridge \\
  advait@microsoft.com \\
  \And
  Andrew D. Gordon \\
  Microsoft Research \\
  University of Edinburgh \\
  adg@microsoft.com \\
  \And
  Carina Negreanu \\
  Microsoft Research \\
  cnegreanu@microsoft.com \\
  \AND
  Christian Poelitz \\
  Microsoft Research \\
  cpoelitz@microsoft.com \\
  \And
  Sruti Srinivasa Ragavan \\
  Microsoft Research \\
  a-srutis@microsoft.com \\
  \And
  Ben Zorn \\
  Microsoft Research \\
  ben.zorn@microsoft.com \\
}

\date{}

\begin{document}
\maketitle

\thispagestyle{plain}
\pagestyle{plain}

\begin{figure}[h]
	\centering
	\includegraphics[width=\textwidth]{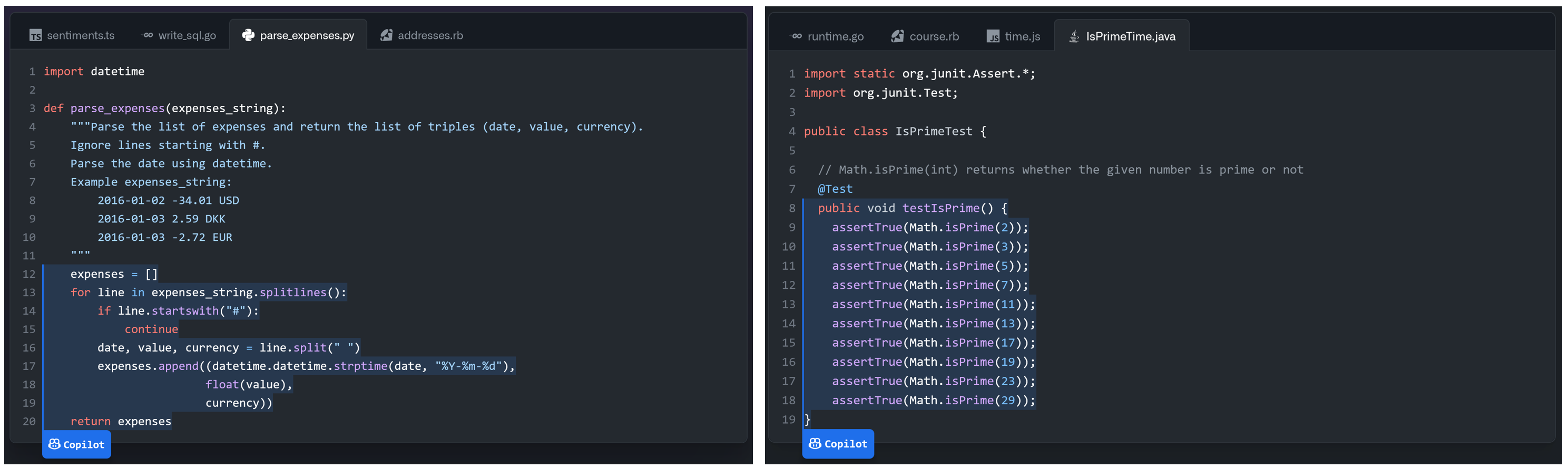}
	\caption{Code generation using the GitHub Copilot editor extension. The portion highlighted in blue has been generated by the model. Left: a function body, generated based on a textual description in a comment. Right: a set of generated test cases. Source: \url{copilot.github.com}}
	\label{fig:copilot1}
\end{figure}

\begin{abstract}
Large language models, such as OpenAI's codex and Deepmind's AlphaCode, can generate code to solve a variety of problems expressed in natural language. This technology has already been commercialised in at least one widely-used programming editor extension: GitHub Copilot.

In this paper, we explore how programming with large language models (LLM-assisted programming) is similar to, and differs from, prior conceptualisations of programmer assistance. We draw upon publicly available experience reports of LLM-assisted programming, as well as prior usability and design studies. We find that while LLM-assisted programming shares some properties of compilation, pair programming, and programming via search and reuse, there are fundamental differences both in the technical possibilities as well as the practical experience. Thus, LLM-assisted programming ought to be viewed as a new way of programming with its own distinct properties and challenges. 

Finally, we draw upon observations from a user study in which non-expert end user programmers use LLM-assisted tools for solving data tasks in spreadsheets. We discuss the issues that might arise, and open research challenges, in applying large language models to end-user programming, particularly with users who have little or no programming expertise.
\end{abstract}

\section{Introduction}
Inferential assistance for programmers has manifested in various forms, such as programming by demonstration, declarative programming languages, and program synthesis (Section~\ref{sec:prior_models}). Large language models such as GPT mark a quantitative and qualitative step-change in the automatic generation of code and natural language text. This can be attributed to cumulative innovations of vector-space word embeddings, the transformer architecture, large text corpora, and pre-trained language models (Section~\ref{sec:large_language_models}).

These models have been commercialised in the form of APIs such as OpenAI Codex, or as programmer-facing tools such as GitHub Copilot and Tabnine. These tools function as a sort of advanced autocomplete, able to synthesize multiple lines of code based on a prompt within the code editor, which may be natural language (e.g., a comment), code (e.g., a function signature) or an ad-hoc mixture. The capabilities of such tools go well beyond traditional syntax-directed autocomplete, and include the ability to synthesize entire function bodies, write test cases, and complete repetitive patterns (Section~\ref{sec:commercial}). These tools have reliability, safety, and security implications (Section~\ref{sec:reliability}).

Prior lab-based and telemetric research on the usability of such tools finds that developers generally appreciate the capabilities of these tools and find them to be a positive asset to the development experience, despite no strong effects on task completion times or correctness. Core usability issues include the challenge of correctly framing prompts as well as the effort required to check and debug generated code (Section~\ref{sec:prior_studies}).

Longitudinal experience reports of developers support some of the lab-based findings, while contradicting others. The challenges of correctly framing prompts and the efforts of debugging also appear here. However, there are many reports that these tools do in fact strongly reduce task time (i.e., speed up the development process) (Section~\ref{sec:experience_reports}).

Programming with large language models invites comparison to related ways of programming, such as search, compilation, and pair programming. While there are indeed similarities with each of these, the empirical reports of the experience of such tools also show crucial differences. Search, compilation, and pair programming are thus found to be inadequate metaphors for the nature of LLM-assisted programming; it is a distinct way of programming with its own unique blend of properties (Section~\ref{sec:metaphors}).

While LLM-assisted programming is currently geared towards expert programmers, arguably the greatest beneficiaries of their abilities will be non-expert end-user programmers. Nonetheless, there are issues with their direct application in end-user programming scenarios. Through a study of LLM-assisted end-user programming in spreadsheets, we uncover issues in intent specification, code correctness, comprehension, LLM tuning, and end-user behaviour, and motivate the need for further study in this area (Section~\ref{sec:eup}).

\section{Prior conceptualisations of intelligent assistance for programmers}
\label{sec:prior_models}

What counts as `intelligent assistance' can be the subject of some debate. Do we select only features that are driven by technologies that the artificial intelligence research community (itself undefined) would recognise as artificial intelligence? Do we include those that use expert-coded heuristics? Systems that make inferences a human might disagree with, or those with the potential for error? Mixed-initiative systems \citep{horvitz1999principles}? Or those that make the user feel intelligent, assisted, or empowered? While this debate is beyond the scope of this paper, we feel that to properly contextualise the qualitative difference made by large language models, a broad and inclusive approach to the term `intelligence' is required.


End-user programming has long been home to inferential, or intelligent assistance. The strategy of direct manipulation \citep{shneiderman1993direct} is highly successful for certain types of limited, albeit useful, computational tasks, where the interface being used (``what you see'', e.g., a text editor or an image editor) to develop an information artefact can represent closely the artefact being developed (``what you get'', e.g., a text document or an image). However, this strategy cannot be straightforwardly applied to programs. Programs notate multiple possible paths of execution simultaneously, and they define ``behaviour to occur at some future time'' \citep{blackwell2002programming}. Rendering multiple futures in the present is a core problem of live programming research \citep{tanimoto2013perspective}, which aims to externalise programs as they are edited \citep{basman2016software}.

The need to bridge the abstraction gap between direct manipulation and multiple paths of execution led to the invention of programming by demonstration (PBD) \citep{kurlander1993watch, lieberman2001your, myers1992demonstrational}. A form of inferential assistance, PBD allows end-user programmers to make concrete demonstrations of desired behaviour that are generalised into executable programs. Despite their promise, PBD systems have not achieved widespread success as end-user programming tools, although their idea survives in vestigial form as various ``macro recording'' tools, and the approach is seeing a resurgence with the growing commercialisation of ``robotic process automation''.


Programming language design has long been concerned with shifting the burden of intelligence between programmer, program, compiler, and user. Programming language compilers, in translating between high-level languages and machine code, are a kind of intelligent assistance for programmers. The declarative language Prolog aspired to bring a kind of intelligence, where the programmer would only be responsible for specifying (``declaring'') \emph{what} to compute, but not \emph{how} to compute it; that responsibility was left to the interpreter. At the same time, the language was designed with intelligent applications in mind. Indeed, it found widespread use within artificial intelligence and computational linguistics research \citep{colmerauer1996birth,rouchy2006aspects}.


Formal verification tools use a specification language, such as Hoare triples \citep{DBLP:journals/cacm/Hoare69}, and writing such specifications can be considered programming at a `higher' level of abstraction. Program synthesis, in particular synthesis through refinement, aims at intelligently transforming these rules into executable and correct code. However, the term ``program synthesis'' is also used more broadly, and programs can be synthesised from other sources than higher-level specifications. Concretely, program synthesis by example, or simply programming by example (PBE), facilitates the generation of executable code from input-output examples. An example of successfully commercialised PBE is Excel's Flash Fill \citep{DBLP:conf/popl/Gulwani11}, which synthesises string transformations in spreadsheets from a small number of examples.


The Cognitive Dimensions framework \citep{green1989cognitive,green1998cognitive} identifies three categories of programming activity: authoring, transcription, and modification. Modern programmer assistance encompasses each of these. For example, program synthesis tools transform the direct authoring of code into the (arguably easier) authoring of examples. Intelligent code completions \citep{marasoiu2015empirical} support the direct authoring of code. Intelligent support for reuse, such as smart code copy/paste \citep{allamanis2017smartpaste} support transcription, and refactoring tools \citep{hermans2015detecting} support modification. Researchers have investigated inferential support for navigating source code \citep{henley2014patchworks}, debugging \citep{williams2020understanding}, and selectively undoing code changes \citep{yoon2015supporting}. Additionally, intelligent tools can also support learning \citep{cao2015idea}.



\cite{DBLP:journals/csur/AllamanisBDS18} review work at the intersection of machine learning, programming languages, and software engineering. They seek to adapt methods first developed for natural language, such as language models, to source code.
The emergence of large bodies of open source code, sometimes called ``big code'', enabled this research area.
Language models are sensitive to lexical features like names, code formatting, and order of methods, while traditional tools like compilers or code verifiers are not.
Through the ``naturalness hypothesis'', which claims that ``software is a form of human communication;
software corpora have similar statistical properties to natural language corpora;
the authors claim that these properties can be exploited to build better software engineering tools.'' Some support for this hypothesis comes from research that used $n$-gram models to build a code completion engine for Java that outperformed Eclipse's completion feature \citep{DBLP:conf/icse/HindleBSGD12,DBLP:journals/cacm/HindleBGS16}.
This approach can underpin recommender systems (such as code autocompletion), debuggers, code analysers (such as type checkers \citep{DBLP:conf/popl/RaychevVK15}), and code synthesizers.
%
%
We can expect the recent expansion in capability of language models, discussed next, to magnify the effectiveness of these applications.

\section{A brief overview of large language models for code generation}
\label{sec:large_language_models}

\subsection{The transformer architecture and big datasets enable large pre-trained models}
In the 2010s, natural language processing has evolved in the development of language models (LMs,) tasks, and evaluation. \cite{NIPS2013_9aa42b31} introduced Word2Vec, where vectors are assigned to words such that similar words are grouped together. It relies on co-occurrences in text (like Wikipedia articles), though simple instantiations ignore the fact that words can have multiple meanings depending on context. Long short-term memory (LSTM) neural networks \citep{10.1162/neco.1997.9.8.1735, 10.5555/2969033.2969173} and later encoder-decoder networks, account for order in an input sequence. Self-attention \citep{10.5555/3295222.3295349} significantly simplified prior networks by replacing each element in the input by a weighted average of the rest of the input. Transformers combined the advantages of (multi-head) attention and word embeddings, enriched with positional encodings (which add order information to the word embeddings) into one architecture. While there are many alternatives to transformers for language modelling, in this paper when we mention a language model we will usually imply a transformer-based language model.


There are large collections of unlabelled text for some widely-spoken natural languages. For example, the Common Crawl project\footnote{\url{https://commoncrawl.org/}} produces around 20 TB of text data (from web pages) monthly. Labelled task-specific data is less prevalent. This makes unsupervised training appealing. Pre-trained LMs \citep{ijcai2021-612} are commonly trained to perform next-word prediction (e.g., GPT~\citep{brown2020language}) or filling a gap in a sequence (e.g., BERT~\citep{devlin-etal-2019-bert}).



Ideally, the ``general knowledge'' learnt by pre-trained LMs can then be transferred to downstream language tasks (where we have less labelled data) such as question answering, fiction generation, text summarisation, etc. Fine-tuning is the process of adapting a given pre-trained LM to different downstream tasks by
introducing additional parameters and training them using task-specific objective functions. In certain cases the pre-training objective also gets adjusted to better suit the downstream task. Instead of (or on top of) fine-tuning, the downstream task can be reformulated to be similar to the original LLM training. In practice, this means expressing the task as a set of instructions to the LLM via a prompt. So the goal, rather than defining a learning objective for a given task, is to find a way to query the LLM to directly predict for the downstream task. This is sometimes referred to as Pre-train, Prompt, Predict.\footnote{\url{http://pretrain.nlpedia.ai/}}

\subsection{Language models tuned for source code generation}

The downstream task of interest to us in this paper is \emph{code generation}, where the input to the model is a mixture of natural language comments and code snippets, and the output is new code. Unlike other downstream tasks, a large corpus of data is available from public code repositories such as GitHub. Code generation can be divided into many sub-tasks, such as variable type generation, e.g. \citep{Wei2020LambdaNetPT}, comment generation, e.g. \citep{liu2021retrievalaugmented}, duplicate detection, e.g \citep{Mou2016ConvolutionalNN}, code migration from one language to another e.g. \citep{Nguyen2015DivideandConquerAF} etc. A recent benchmark that covers many tasks is CodeXGLUE \citep{Lu2021CodeXGLUEAM}.


LLM technology has brought us within reach of full-solution generation. Codex \citep{Chen2021EvaluatingLL}, a version of GPT-3 fine-tuned for code generation, can solve on average 47/164 problems in the HumanEval code generation benchmark, in one attempt. HumanEval is a set of 164 hand-written programming problems, which include a function signature, docstring, body, and several unit tests, with an average of 7.7 tests per problem. Smaller models  have followed Codex, like GPT-J\footnote{\url{https://huggingface.co/docs/transformers/main/model\_doc/gptj}} (fine-tuned on top of GPT-2), CodeParrot\footnote{\url{https://huggingface.co/blog/codeparrot}} (also fine-tuned on top of GPT-2, targets Python generations), PolyCoder \citep{Xu2022ASE}(GPT-2 style but trained directly on code).

LLMs comparable in size to Codex include AlphaCode~\citep{Li2022CompetitionLevelCG} and PaLM-Coder~\citep{Chowdhery2022PaLMSL}. AlphaCode is trained directly on GitHub data and fine-tuned on coding competition problems. It introduces a method to reduce from a large number of potential solutions (up to millions) to a handful of candidates (competitions permit a maximum of 10). On a dataset of 10000 programming problems, Codex solves around 3\% of the problems within 5 attempts, versus AlphaCode which solves 4-7\%. In competitions for which it was fine-tuned (CodeContests) AlphaCode achieves a 34\% success rate, on par with the average human competitor.


Despite promising results there are known shortcomings. Models can directly copy full solutions or key parts of the solutions from the training data, rather than generating new code. Though developers make efforts to clean and retain only high-quality code, there are no guarantees of correctness and errors can be directly propagated through generations.

Codex can also produce syntactically incorrect or undefined code, and can invoke functions, variables, and attributes that are undefined or out of scope. Moreover, Codex struggles to parse through increasingly long and higher-level or system-level specifications which can lead to mistakes in binding operations to variables, especially when the number of operations and variables in the docstring is large. Various approaches have been explored to filter out bad generations or repair them, especially for syntax errors.

Consistency is another issue. There is a trade-off between non-determinism and generation diversity. Some parameter settings can control the diversity of generation (i.e., how diverse the different generations for a single prompt might be), but there is no guarantee that we will get the same generation if we run the system at different times under the same settings. To alleviate this issue in measurements, metrics such as \texttt{pass@k} (have a solution that passes the tests within $k$ tries) have been modified to be probabilistic.




\section{Commercial programming tools that use large language models}
\label{sec:commercial}

\begin{figure}[t]
	\centering
	\includegraphics[width=0.9\textwidth]{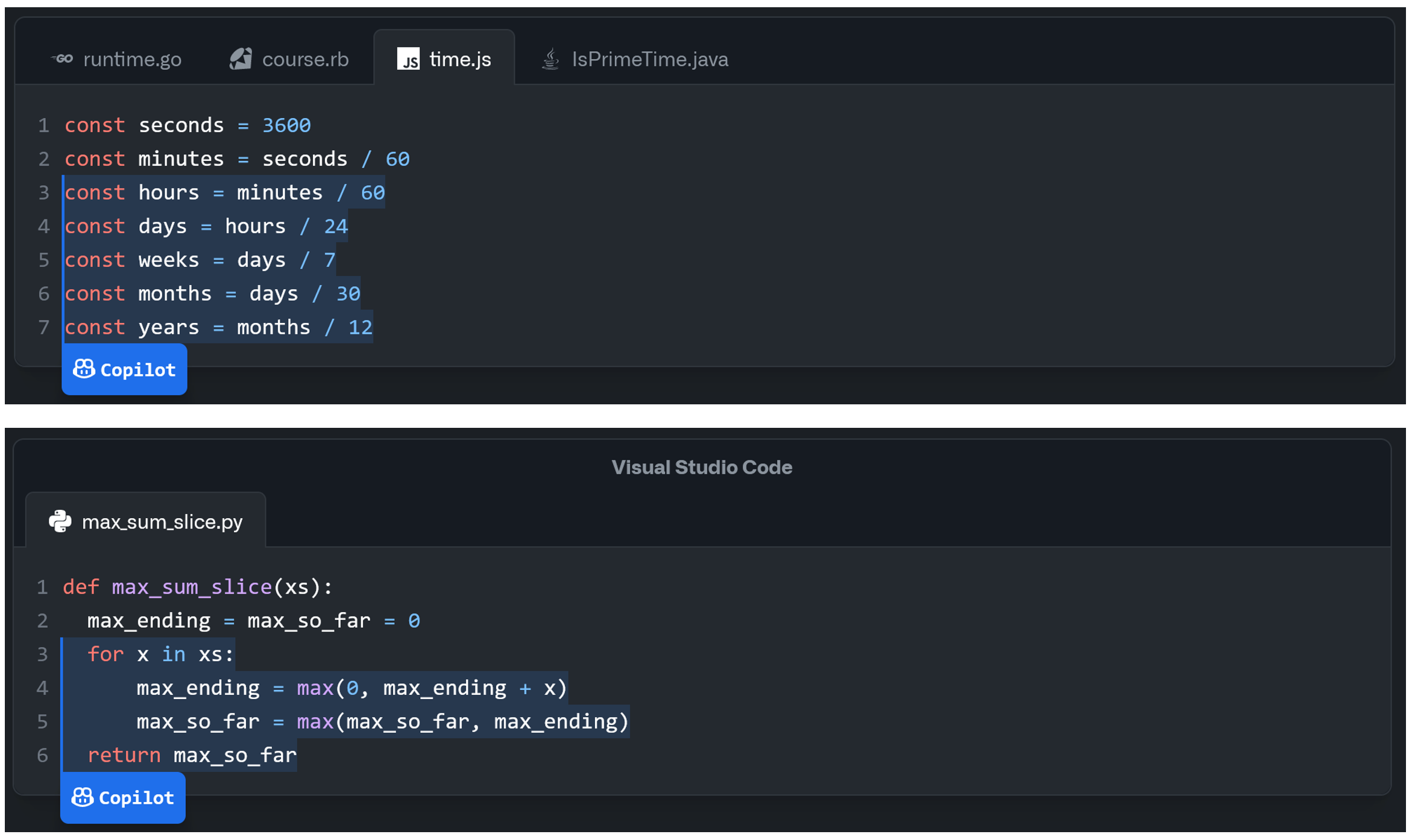}
	\caption{Code generation with GitHub Copilot. The portion highlighted in blue has been generated by the model. Above: a pattern, extrapolated based on two examples. Below: a function body, generated from the signature and the first line. Source: \url{copilot.github.com}}
	\label{fig:copilot2}
\end{figure}

\begin{figure}[t]
	\centering
	\includegraphics[width=0.7\textwidth]{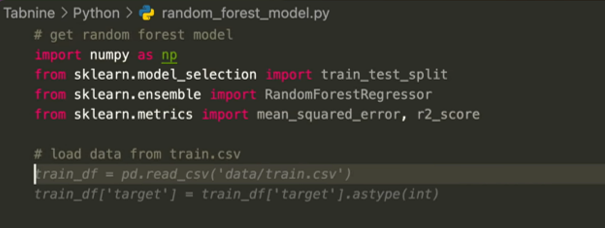}
	\caption{Code generation using the Tabnine editor extension. The grey text after the cursor is being suggested by the model based on the comment on the preceding line. Source: \url{tabnine.com}}
	\label{fig:tabnine}
\end{figure}

\begin{figure}[h]
	\centering
	\includegraphics[width=0.9\textwidth]{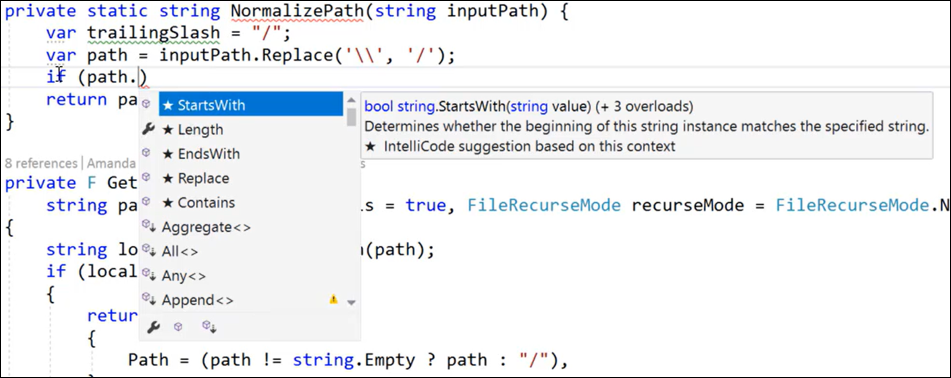}
	\caption{API suggestion using the Visual Studio IntelliCode feature. Source: \cite{silver_2018}}
	\label{fig:intellicode}
\end{figure}

OpenAI Codex is a version of GPT that is fine-tuned on publicly available source code \citep{chen2021codex}. While Codex itself is not a programmer-facing tool, OpenAI has commercialised it in the form of an API that can be built upon.

The principal commercial implementation of Codex thus far has been in Github Copilot.\footnote{\url{https://copilot.github.com/}} Copilot is an extension that can be installed into code editors such as Neovim, Jetbrains, and Visual Studio Code. Copilot uses Codex, drawing upon the contents of the file being edited, related files in the project, and file paths or URLs of repositories. When triggered, it generates code at the cursor location, in much the same way as autocomplete.

To help expand developer expectations for the capabilities of Copilot beyond the previous standard uses of autocomplete, suggested usage idioms for Copilot include: writing a comment explaining what a function does, and the function signature, and allowing Copilot to complete the function body; completing boilerplate code; and defining test cases (Figures~\ref{fig:copilot1}~and~\ref{fig:copilot2}). Programmers can cycle between different generations from the model, and once a particular completion has been accepted it can be edited like any other code.

As of 23 June 2022, Amazon has announced a Copilot-like feature called CodeWhisperer,\footnote{\url{https://aws.amazon.com/codewhisperer/features/}} which also applies a large language model trained on a corpus of source code to generate autocompletions based on comments and code. The marketing material describes a set of safety features, such as: detecting when generated code is similar to code in the training set, detecting known security vulnerabilities in the generated code, and \emph{``removing code that may be considered biased and unfair''} (although this latter claim induces skepticism). At present CodeWhisperer is not widely available and thus little is known of its use in practice.

Other commercial implementations of AI-assisted autocompletion features include Visual Studio Intellicode \citep{silver_2018} (Figure~\ref{fig:intellicode}) and Tabnine (Figure~\ref{fig:tabnine})\footnote{\url{https://www.tabnine.com/}}. These are more limited in scope than Copilot and their user experience is commensurable to that of using `traditional' autocomplete, i.e., autocomplete that is driven by static analysis, syntax, and heuristics.\footnote{As of 15 June 2022, Tabnine has announced a shift to language model-driven autocompletion that more closely resembles the abilities of Copilot \citep{weiss_2022}.}.The structure of the machine learning model used by these implementations is not publicly disclosed; however, both rely on models that have been trained on large corpora of publicly available source code.

It is interesting to note, that despite the wide variety of types of intelligent programmer assistance we have discussed in Section~\ref{sec:prior_models} for several aspects of programming (authoring, transcription, modification, debugging, and learning), commercial implementations of assistance based on large language models thus far are aimed primarily at authoring. Authoring can be viewed as the first natural application of a generative language model, but the programming knowledge in these models can of course be used for assisting programmers in other activities, too.


\section{Reliability, safety, and security implications of code-generating AI models}
\label{sec:reliability}

AI models that generate code present significant challenges to issues related to reliability, safety, and security.  Since the output of the model can be a complex software artifact, determining if the output is ``correct'' needs a much more nuanced evaluation than simple classification tasks. Humans have trouble evaluating the quality of software, and practices such as code review, applying static and dynamic analysis techniques, etc., have proven necessary to ensure good quality of human-written code. Current methods for evaluating the quality of AI-generated code, as embodied in benchmarks such as HumanEval~\citep{chen2021codex}, MBPP~\citep{austin2021mbpp}, and CodeContests~\citep{li2022:alphacode}, determine functional correctness of entire functions based on a set of unit tests. Such evaluation approaches fail to consider issues of code readability, completeness, or the presence of potential errors that software developers constantly struggle to overcome. 

Previous work~\citep{chen2021codex} explores numerous implications of AI models that generate code, including issues of over-reliance, misalignment (the mismatch between what the user prompt requests and what the user really wants), bias, economic impact, and security implications.  While these topics each are extensive and important, due to space limitations we only briefly mention them here and point to additional related work when possible.  Over-reliance occurs when individuals make optimistic assumptions about the correctness of the output of an AI model, leading to harm. For code generating models, users may assume the code is correct, has no security vulnerabilities, etc. and those assumptions may lead to lower quality or insecure code being written and deployed.  Existing deployments of AI models for code, such as GitHub Copilot~\citep{ziegler_2021}, have documentation that stresses the need to carefully review, test, and vet generated code just as a developer would vet code from any external source.  It remains to be seen if over-reliance issues related to AI code generation will result in new software quality challenges.

Since AI that generates code is trained on large public repositories, there is potential for low-quality training data to influence models to suggest low-quality code or code that contains security vulnerabilities.  One early study of GitHub Copilot~\citep{pearce2021copilotsecurity} examines whether code suggestions may contain known security vulnerabilities in a range of scenarios and finds cases where insecure code is generated.  Beyond carefully screening new code using existing static and dynamic tools that detect security vulnerabilities in human-generated code, there are also possible mitigations that can reduce the likelihood that the model will make such suggestions.  These include improving the overall quality of the training data by removing low-quality repositories, and fine-tuning the large-language model specifically to reduce the output of known insecure patterns.

\section{Usability and design studies of AI-assisted programming}
\label{sec:prior_studies}

\citet{vaithilingam2022expectation} conducted a within-subjects comparative study (n=24) of Github Copilot, comparing its user experience to that of traditional autocomplete (specifically, the \emph{Intellisense} plugin, not the same as the \emph{Intellicode} feature mentioned previously). Participants failed to complete the tasks more often with Copilot than with Intellisense, and there was no significant effect on task completion time. Perhaps unsurprisingly, the authors find that assessing the correctness of generated code is difficult and an efficiency bottleneck, particularly when the code generated has a fundamental flaw or inefficiency that leads the programmer on an ultimately unsuccessful `wild goose chase' of repair or debugging. However, the overwhelming majority (19 of 24) of participants reported a strong preference for Copilot in a post-task survey. While participants were less confident about the code generated by Copilot, they almost universally (23 of 24) perceived it as more helpful, because it had the potential for generating useful starting points and saving the programmer the effort of searching online for documented solutions that could be the basis for reuse.

\citet{ziegler2022productivity} conducted a survey (n=2,047) of the perceived productivity of Copilot users in the USA. They matched these to telemetric usage measurements of the Copilot add-in, which included metrics such as how often an auto-completion was shown, how often it was accepted, how often it persisted unchanged in the document for a certain time period, how often it persisted with minor variations (e.g., measured by Levenshtein distance) and so on. They find that the acceptance rate (the ratio of accepted suggestions to shown suggestions) is the strongest predictor of users' perceived productivity due to Copilot. Fascinatingly, they find that the pattern of acceptance rates for all users in aggregate follows a daily and weekly ``circadian'' rhythm, such that users are more likely to accept Copilot completions out of working-hours and on weekends. However, for any given user, the acceptance rate depends on that user's normal working hours; suggestions outside of normal working hours are less likely to be accepted. Future work is needed to see whether this finding replicates, and if so to establish how and why acceptance rates are so significantly affected by working hours.

\citet{xu2022ide} conducted a within-subjects study (n=31) comparing the programming experience with and without a code generation plugin. Their experimental plugin takes the form of a text field in which the user enters a natural language prompt, the system responds with a list of code snippets, and when clicked the desired snippet is inserted at the cursor. This workflow differs from Copilot's, where the `prompt' is text within the source file, and can contain a mix of natural language comments and code. The plugin supported both code generation (using a tree-based neural network) and code snippet retrieval (searching the programming forum Stack Overflow). Results from both generation and retrieval are shown in the same list, but visually demarcated. The authors found no significant effect of the plugin on task completion time or program correctness. They found that simple queries were more likely to be answered correctly through generation, and more complex queries requiring multiple steps were more likely to be answered correctly though retrieval, and that it was possible to predict which approach would succeed based on the word content of the queries. Further, they found that most (60\%) natural language queries that participants wrote in their experiment were not sufficiently well-specified for a human expert to write code implementing those intents. Retrieved snippets were edited more often than generated snippets, mostly to rename identifiers and choose different parameters. In a post-experiment survey, participants reported mostly feeling neutral or somewhat positive (30 of 31). These participants felt that the plugin was helpful for finding snippets they were aware of but cannot recall, and less disruptive than using a browser, but the interaction worked better when the developer had a pre-existing knowledge of the target APIs and frameworks, and it took experimentation to understand the ``correct way'' to formulate queries. There was no clear indication of preference between retrieval and generation.

\cite{jiang2022discovering} developed an LLM-based tool for converting natural language statements to code. As in \cite{xu2022ide}, prompts are entered in a pop-up dialog invoked at the cursor from within a code editor, rather than as comments. In a study $(n=14)$, participants were given a week to complete two website-building tasks with the tool, while recording the screen, and were interviewed afterwards. As in other studies, participants saw utility in the tool for facilitating quick API lookups and for writing boilerplate code. They found that novice programmers' queries were mainly natural language, whereas experts were more likely to mix code into their requests. While some queries were abstract, and expressed high-level goals, most had low granularity, being ``roughly equivalent to a line of code''. To cope with model failures, participants used a variety of strategies to reword their query, such as reducing the scope of the request or replacing words with alternatives, but no particular strategy was observed to be more effective than any other. Participants struggled with forming a mental model of what the model can understand and the ``syntax'' of the language it required -- this is precisely the \emph{fuzzy abstraction matching} problem we described earlier, which the authors call an ``uncanny valley''.  The authors suggest possible solutions such as automated rewording of prompts, suggesting simpler tasks, suggesting task breakdowns, and better onboarding and tutorials.

\cite{barke2022grounded} studied how programmers ($n=20$) use GitHub Copilot to complete short programming tasks in Python, Rust, Haskell, and Java. Through analysis of screen recordings, the authors identifed two primary modes of interaction with Copilot: \emph{acceleration}, where the programmer has a well-formed intent and Copilot speeds up code authoring in ``small logical units'', and \emph{exploration}, where Copilot suggestions are used to assist the planning process, ``help them get started, suggest potentially useful structure and API calls, or explore alternative solutions''. In acceleration, long code suggestions, which take time to read and evaluate, can break the programmer's flow. Participants developed heuristics for quickly scanning suggestions, such as looking for the presence of certain keywords. In exploration, participants were more likely to prompt using purely natural language comments, rather than a mix of comments and code. Moreover, these prompt comments were often `cleaned' subsequent to accepting a suggestion, which implies a form of `instruction language' that is separate from `explanation language'.

\cite{madi2022readable} compared the readability of code generated by Copilot with that of code written by human programmers in a user study ($n=21$). They found that model generated code is comparable in complexity and readability to human-authored code.

\begin{figure}[t]
	\centering
	\fbox{\includegraphics[width=\textwidth]{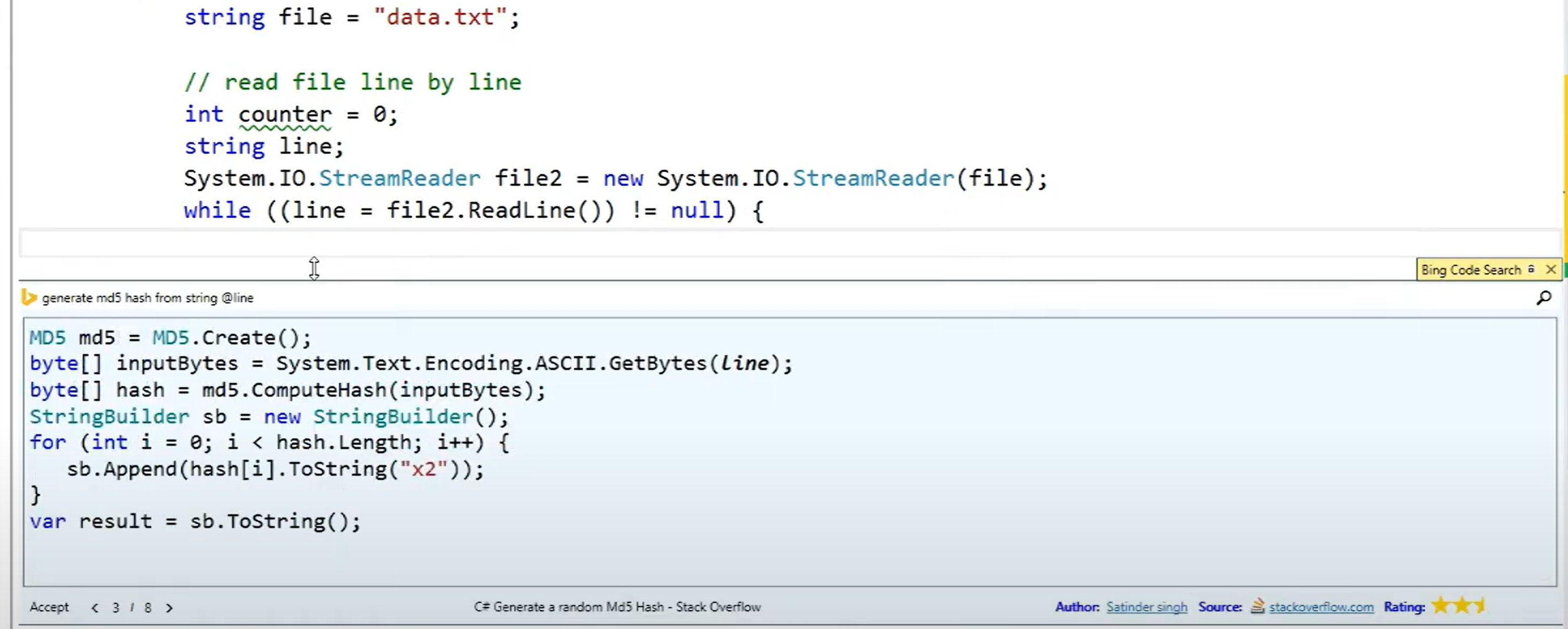}}
	\caption{Searching for code snippets using Bing Developer Assistant. A result for Stack Overflow is shown. Note how the query ``\texttt{generate md5 hash from string @line}'' contains a hint about the identifier \texttt{line}, which is used to rewrite the retrieved snippet. Source: \url{https://www.microsoft.com/en-us/research/publication/building-bing-developer-assistant/}}
	\label{fig:copilot2}
\end{figure}

The Bing Developer Assistant \citep{wei2015building, zhang2016bing} (also referred to as Bing Code Search) was an experimental extension for Visual Studio initially released in 2015.
It enabled an in-IDE, identifier-aware search for code snippets from forums such as Stack Overflow. It had the ability to rewrite retrieved code to use identifiers from the programmer's current file. A user study (n=14) comparing task time in performing 45 short programming tasks with the extension versus regular web search found on average 28\% of time was saved with the extension. Morever telemetry data gathered over three weeks (representing around 20,000 users and around 3,000 queries per day) showed that several programmers used the feature frequently. Some used it repeatedly for related problems in quick succession, showing its use in multi-step problems. Others issued the same query multiple times on separate days, suggesting that the speed of auto-completion was useful even if the programmer knew the solution.


\input{sec_experience_reports}
\input{sec_related_models}

\section{Issues with application to end-user programming}
\label{sec:eup}

The benefits and challenges of programming with LLMs discussed so far concern the professional programmer, or a novice programmer in training. They have formal training in programming and, often, some understanding of the imperfect nature of AI-generated code. But the majority of people who program do not fall into this category. Instead, they are ordinary end users of computers who program to an end. Such end-user programmers often lack knowledge of programming, or the workings of AI. They also lack the inclination to acquire those skills. 

It is reasonable to say that such end-user programmers (e.g., accountants, journalists, scientists, business owners) stand to benefit the most from AI assistance, such as LLMs. In one ideal world, an end-user wanting to accomplish a task could do so by simply specifying their intent in familiar natural language  without prior knowledge of the underlying programming model, or its syntax and semantics. The code will get generated and even automatically run to produce the desired output.

However, as we have seen so far, the world is not ideal and even trained programmers face various challenges when programming with AI. These challenges are only exacerbated for end-user programmers, as a study by \cite{ragavan2022gridbook} observes.

\begin{figure}[t]
	\centering
	\fbox{\epsfig{file=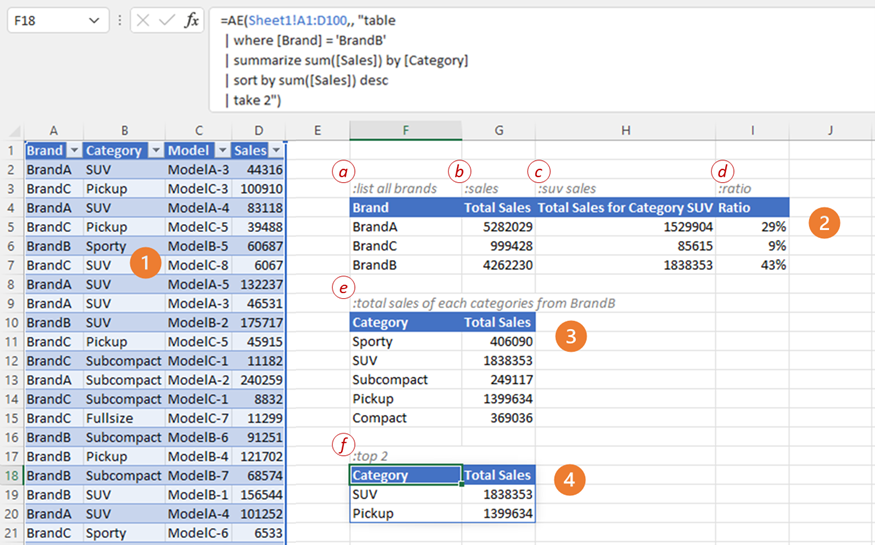, scale=1}}
	\caption{GridBook interface showing natural language formula in the spreadsheet grid.}
	\label{fig:gridbook}
\end{figure}

Participants in the study were data analysts (n=20) conducting exploratory data analysis in GridBook, a natural-language augmented spreadsheet system. In GridBook (Figure \ref{fig:gridbook}, adopted from \cite{ragavan2022gridbook}) users can write spreadsheet formulas using the natural language (Figure \ref{fig:gridbook}: a-f); a formal formula is then synthesized from the natural language utterance. GridBook also infers the context of an utterance; for example, in Figure \ref{fig:gridbook}, the query in label 4 is a follow-up from label 3.  Both the natural language utterance and the synthesized formula are persisted for users to edit and manipulate. 

\subsection{Issue 1: Intent specification, problem decomposition and computational thinking} When attempting to accomplish data analysis tasks using natural language, participants had to refine their specification of intent in the natural language several times, before they arrived at the desired result (if they did). The NL utterances were often underspecified, ambiguous, too complex, or contained domain phrases not specified in the context (e.g., in the data being analyzed). Thus, the first issue is to communicate the capabilities of the system, and make it interpretable so users can see how their prompt is being interpreted. 

End-user programmers often lack computational thinking skills \citep{wing2011computationalthinking}, such as the ability to decompose problems into subproblems, reformulate problems in ways that can be computed by a system, etc. However, effective use of LLMs such as Codex requires such skills. For example, if these models are most accurate when solutions to a problem are single line, then the user should be able to break their problem into smaller sub-problems each of which can be solved in one or two lines. Moreover, they might also lack the ability to frame a problem as generic computational problems, rather than domain-specific problems. For example, a realtor is more likely to ask ``which is the largest house'' (declaratively), instead of ``which is the house with maximum constructed area'' (procedurally).

Therefore, end-user computing environments powered by AI should help end-user programmers think ``computationally'': they must aid users in breaking down their problems to smaller steps, or guiding users towards alternative strategies to specify or solve a problem (e.g., providing examples, offering alternatives) or even seek procedural prompts where needed (e.g., for disambiguation).

\subsection{Issue 2: Code correctness, quality and (over)confidence} The second challenge is in verifying whether the code generated by the model is correct. In GridBook, users were able to see the natural language utterance, synthesized formula and the result of the formula. Of these, participants heavily relied on `eyeballing' the final output as a means of evaluating the correctness of the code, rather than, for example, reading code or testing rigorously.

While this lack of rigorous testing by end-user programmers is unsurprising, some users, particularly those with low computer self-efficacy, might overestimate the accuracy of the AI, deepening the overconfidence end-user programmers are known to have in their programs' accuracy \citep{panko2008reducing}. Moreover, end-user programmers might not be able to discern the quality of non-functional aspects of the generated code, such as security, robustness or performance issues.

\subsection{Issue 3: Code comprehension and maintenance} A third challenge with AI-driven programming is the issue of code comprehension. During GridBook's user evaluation, participants mentioned that the generated formulas are hard to understand, even when users were familiar with the target language. This has potentially severe consequences: from evaluating the accuracy of the program by verifying logic, to the ability to customize code, to future debugging and reuse. As we discussed earlier, this problem also exists for trained developers. 

One approach to address this issue is for the AI system to include some notion of code readability or comprehensibility as a factor in code synthesis, such as during the learning phase, or when ranking suggestions, or even take it as input to the model (similar to the `temperature' parameter in Codex). This approach is useful more broadly to synthesize high quality code, such as optimizing for performance or robustness. A second solution to tackle the comprehension problem is to explain the generated code to their users in a manner that is less `programmerese' and more centered around the user's current task and context. Initial evidence suggests that participants were open to these ideas; thus, these areas are ripe for future exploration.


\subsection{Issue 4: Consequences of automation in end-user programming}
In any AI system, we need to consider the consequences of automation. End-user programmers are known to turn to local experts or gardeners (end-user programmers with interest and expertise in programming who serve as gurus in the end-user programming environment) when they are unable to solve a part of the problem \citep{nardiEUPbook, sarkar2018people}. Task-orientation tendencies combined with challenges of completing their tasks easily also leaves end-user programmers with limited attention for testing, or carefully learning what is going on with their programs. Assuming that LLMs and associated user experiences will improve in the coming years, making end-user programming faster with LLMs than without, it is tempting to wonder whether the programmer can be persuaded to invest the saved time and attention to aspects such as learning or testing their programs; if so, what would it take to influence behaviour changes?

Another question is in the role of such experts. We conjecture that LLMs or similar AI capabilities will soon be able to answer a sizeable fraction of questions that end-user programmers will go to local experts for. An open question therefore is how the ecosystem of end-user programmers in organizations will change in their roles, importance and specialities. For example, will gardeners take on the role of educating users on better taking advantage of AI? If so, how can we communicate the working of such AI systems to technophile users and early adopters, so they can enable others in the organization?

\subsection{Issue 5: No code, and the dilemma of the direct answer}
Finally, it is not a foregone conclusion that users are even interested in code. As Blackwell's model of attention investment notes, in many cases the user may be content to perform an action manually, rather than invest in creating a reusable automation \citep{blackwell2002first,williams2020understanding}. Spreadsheet users, in particular, are often not sensitive to the level of automation or automatability of a given workflow, using a mix of manual, automated, and semi-automated techniques to achieve the goal at hand \citep{pandita2018no}. 

Spreadsheet users often need ad-hoc transformations of their data that they will, in all likelihood, never need again. It may be that we can express this transformation as a program, but if the user is interested in the output and not the program, is it important, or even necessary, to communicate this fact to the user? One can argue that increasing the user's awareness of the flexibility and fallibility of the process of delivering an inferred result (i.e., enabling them to \emph{critically evaluate} the output \citep{sarkar2015interactive}) can build agency, confidence, trust, and resilience. This issue is related to information retrieval's ``dilemma of the direct answer'' \citep{potthast2021dilemma}, raised in response to the increased phenomenon of search engines directly answering queries in addition to simply listing retrieved results. 

However, if the programming language used is not related to the languages familiar to the end-user, or the user is a complete novice, it is exceedingly difficult for them to make any sense of it, as was shown by \cite{lau2021tweakit} in their study of Excel users encountering Python code. Yet, there are socio-technical motivations for using an unfamiliar target language: long-term testing of LLM assistance shows that it shines when paired with high-level APIs that capture use cases well (Section~\ref{sec:experience_reports}). One advantage of the Python ecosystem is that it has an unparalleled set of libraries and APIs for data wrangling. An LLM-assisted tool that emits Excel formulas is therefore less likely to solve user problems than Python statements. In the longer term, this might be mitigated by developing a rich set of data manipulation libraries in the Excel formula language.









\section{Conclusion}
\label{sec:conclusion}

Large language models have initiated a significant change in the scope and quality of program code that can be automatically generated, compared to previous approaches. Experience with commercially available tools built on these models suggests that a they represent a new way of programming. LLM assistance transforms almost every aspect of the experience of programming, including planning, authoring, reuse, modification, comprehension, and debugging. 

In some aspects, LLM assistance resembles a highly intelligent and flexible compiler, or a partner in pair programming, or a seamless search-and-reuse feature. Yet in other aspects, LLM-assisted programming has a flavour all of its own, which presents new challenges and opportunities for human-centric programming research. Moreover, there are even greater challenges in helping non-expert end users benefit from such tools.

\clearpage
\input{experience_report_sources}


\clearpage
\bibliography{references}
\bibliographystyle{apacite}

\end{document}

%% file: sec_experience_reports.tex
\section{Experience reports}
\label{sec:experience_reports}

At present, there is not a lot of research on the user experience of programming with large language models beyond the studies we have summarised in Section~\ref{sec:prior_studies}. However, as the availability of such tools increases, professional programmers will gain long-term experience in their use. Many such programmers write about their experiences on personal blogs, which are then discussed in online communities such as Hacker News. Inspired by the potential for these sources to provide rich qualitative data, as pointed out by Barik \citep{barik2015heart, sarkar2022lambdas}, we draw upon a few such experience reports. A full list of sources is provided Appendix~\ref{apx:experience_report_sources}; below we summarise their key points.

\subsection{Writing effective prompts is hard}
As with several other applications of generative models, a key issue is the writing of prompts that increase the likelihood of successful code generation. The mapping that these models learn between natural language and code is very poorly understood. Through experimentation, some have developed heuristics for prompts that improve the quality of the code generated by the model. One developer, after building several applications and games with OpenAI's \texttt{code-davinci} model (the second generation Codex model), advises to \emph{``number your instructions''} and creating \emph{``logic first''} before UI elements. Another, in using Copilot to build a classifier for natural language statements, suggests to provide \emph{``more detail''} in response to a failure to generate correct code. For example, when asking Copilot to \emph{``binarize''} an array fails, they re-write the prompt to \emph{``turn it into an array where [the first value] is 1 and [the second value] is 0''} -- effectively pseudocode -- which generates a correct result. 

Commenters on Hacker News are divided on the merits of efforts invested in developing techniques for prompting. While some see it as a new level of abstraction for programming, others see it as indirectly approaching more fundamental issues that ought to be solved with better tooling, documentation, and language design:

\begin{quote}
    ``You're not coding directly in the language, but now you're coding in an implicit language provided by Copilot. \el{From what I've seen on Copilot, although it is an impressive piece of tech,} all it really points out is that code documentation and discovery is terrible. But I'm not for sure writing implicit code in comments is really a better approach than seeking ways to make discovery of language and library features more discoverable.''
\end{quote}

\begin{quote}
    ``\el{Notice that} the comments used to generate the code via GitHub Copilot are just another very inefficient programming language.''
\end{quote}

\begin{quote}
    ``[Responding to above] There is nonetheless something extremely valuable about being able to write at different levels of abstraction when developing code. Copilot lets you do that in a way that is way beyond what a normal programming language would let you do, which of course has its own, very rigid, abstractions. For some parts of the code you'll want to dive in and write every single line in painstaking detail. For others \el{`# give me the industry standard analysis of this dataset`} [Copilot] is maybe enough for your purposes. And being able to have that ability, even if you think of it as just another programming language in itself, is huge.''
\end{quote}

Being indiscriminately trained on a corpus containing code of varying ages and (subjective) quality has drawbacks; developers encounter generated code which is technically correct, but contains practices considered poor such as unrolled loops and hardcoded constants. One Copilot user found that:
\begin{quote}
    ``Copilot \el{is a system that's optimized for write. It} has made my code more verbose. Lines of code can be liabilities. Longer files to parse, and more instances to refactor. Before, where I might have tried to consolidate an API surface, I find myself maintaining [multiple instances].''
\end{quote}

Another Copilot user reflected on their experience of trying to generate code that uses the \texttt{fastai} API, which frequently changes: 
\begin{quote}
``\el{} since the latest version of fastai was only released in August 2020, GitHub Copilot was not able to provide any relevant suggestions and instead provided code for using older versions of fastai. \el{} To me, this is a major concern \el{regarding the current usability of GitHub Copilot 4.} If we are using cutting edge tools \el{like PyTorch XLA, JAX, fastai, timm, GitHub} Copilot has no knowledge of this and cannot provide useful suggestions.''
\end{quote}

On the other hand, developers can also be exposed to \emph{better} practices and APIs through these models. The developer that found Copilot to make their code more verbose also observed that:

\begin{quote}
``Copilot gives structure to Go errors . \el{If you program in Go, you know that developers spend a lot of time error handling.} A common idiom is to wrap your errors with a context string [which can be written in an inconsistent, ad-hoc style] \el{, so you can get information about the call stack, something like

return errors.Wrapf(err, "open: \%s", filename)
The problem with any string like that is that it lacks consistency - developers can write whatever they want. There's not a clear strategy either – functions can return errors at multiple points, so you shouldn't just use the parent function name. And sometimes the errors don't come from a function call, they can come from a map lookup or type cast.

} Since using Copilot, I haven't written a single one of these error handling lines manually. On top of that, the suggestions follow a reasonable structure where I didn't know structure had existed before. Copilot showed me how to add structure in my code in unlikely places.
For writing SQL, it helped me write those annoying foreign key names in a consistent format \el{like fk_users_to_teams_user_id.

Discovering new APIs.

} 

[Additionally,] One of the more surprising features has been [that] \el{discovering new API methods, especially for popular libraries. Before, I would context switch between the API documentation and my code. Otherwise, I'd work from memory. Now} I find myself discovering new API methods, either higher-level ones or ones that are better for my use case.''
\end{quote}

In order to discover new APIs, of course, the APIs themselves need to be well-designed. Indeed, in some cases the spectacular utility of large language models can be largely attributed to the fact that API designers have already done the hard work of creating an abstraction that is a good fit for real use cases \citep{myers2016improving,piccioni2013empirical,macvean2016api}. As a developer who used Copilot to develop a sentiment classifier for Twitter posts matching certain keywords remarks, \emph{``These kinds of things are possible not just because of co pilot [sic] but also because we have awesome libraries which have abstracted a lot of tough stuff.''} This suggests that API design, not just for human developers but also as a target for large language models, will be important in the near and mid-term future.

Moreover, breaking down a prompt at the `correct' level of detail is also emerging as an important developer skill. This requires at least some familiarity, or a good intuition, for the APIs available. Breaking down prompts into steps so detailed that the programmer is effectively writing pseudocode, can be viewed as an anti-pattern, and can give rise to the objections cited earlier that programming via large language models is simply a \emph{``very inefficient programming language''}. We term this the problem of \emph{fuzzy abstraction matching}. The problem of figuring out what the system can and can't do, and matching one's intent and instructions with the capabilities of the system, is not new -- it has been well-documented in natural language interaction \citep{mu2019we, luger2016like}. It is also observed in programming notation design as the `match-mismatch' hypothesis \citep{green1992visual, chalhoub2022freedom}. In the broadest sense, these can be seen as special cases of Norman's ``gulf of execution'' \citep{DBLP:journals/hhci/HutchinsHN85}, perhaps the central disciplinary problem of first and second-wave \citep{DBLP:journals/interactions/Bodker15} human-computer interaction research: `how do I get the computer to do what I want it to do?'.


What distinguishes fuzzy abstraction matching from previous incarnations of this problem is the resilience to, and accommodation of, various levels of abstraction afforded by large language models. In previous natural language interfaces, or programming languages, the user needed to form an extremely specific mental model before they could express their ideas in machine terms. In contrast, large language models can generate plausible and correct results for statements at an extremely wide range of abstraction. In the context of programming assistance, this can range from asking the model to write programs based on vague and underspecified statements, requiring domain knowledge to solve, through to extremely specific and detailed instructions that are effectively pseudocode. This flexibility is ultimately a double-edged sword: it has a lower floor for users to start getting usable results, but a higher ceiling for getting users to maximum productivity.


In the context of programming activities, \emph{exploratory programming}, where the goal is unknown or ill-defined \citep{kery2017exploring, sarkar2016phd}, does not fit the framing of fuzzy abstraction matching (or indeed any of the variations of the gulf of execution problem). When the very notion of a crystallised user \emph{intent} is questioned, or when the design objective is for the system to influence the intent of the user (as with much designerly and third-wave HCI work), the fundamental interaction questions change. One obvious role the system can play in these scenarios is to help users refine their own concepts \citep{kulesza2014structured} and decide what avenues to explore. Beyond noting that such activities exist, and fall outside the framework we have proposed here, we will not explore them in greater detail in this paper.


\subsection{The activity of programming shifts towards checking and unfamiliar debugging}

When code can be generated quickly, as observed with the studies in Section~\ref{sec:prior_studies}, checking the correctness of generating code becomes a major bottleneck. This shift, or tradeoff, of faster authoring at the expense of greater time spent checking code, is not without criticism. For some it is the wrong balance of priorities between system and programmer.

Correspondingly, some users have developed heuristics for when the cost of evaluating the correctness of the code is greater than the time or effort saved by code generation, such as to focus on very short (e.g., single line) completions and ignore longer completions. 

Furthermore, some users have found that rather than having suggestions show all the time, which can be distracting and time consuming, more intentional use can be made of Copilot by switching off auto-suggestion and only triggering code completion manually using a keyboard shortcut. However, this requires users to form a mental model of when Copilot is likely to help them in their workflow. This mental model takes time and intentionality to build, and may be incorrect. Moreover, it introduces a new cognitive burden of constantly evaluating whether the current situation would benefit from LLM assistance. Commenters on Hacker News raise these issues:

\begin{quote}
    ``I find I spend my time reviewing Copilot suggestions (which are mostly wrong) rather than thinking about code and actually doing the work.''
\end{quote}

\begin{quote}
    ``\el{I don't think you've actually used it tbh.} It's much quicker to read code than to write it. In addition, 95\% of Copilots suggestions are a single line and they're almost always right (and also totally optional).\el{
    Here's how it actually works in practice
    
    1. Start a line to do an obvious piece of code that is slightly tedious to write 2. Type 2 characters and Copilot usually guesses what you want based on context. Perhaps this time it's off. 3. No matter, just type another 3 characters or so and Copilot catches up and gives a different suggestion. I just hit "tab" and the line is complete
    
    It really shines in writing boiler plate.} I admit that I'm paranoid every time it suggests more than 2 lines so I usually avoid it. \el{But in ~year of using it} I've run into Copilot induced headaches twice. Once was in the first week or so of using it. I sweared off [sic] of using it for anything more than a line then. Eventually I started to ease up since it was accurate so often and then I learned my second lesson with another mistake. \el{Other than that it's done nothing but save me time. It's also been magnificent for learning new languages. I'll usually look up it's suggestions to understand better but even knowing what to look up is a huge service it provides}''
\end{quote}

\begin{quote}
    ``\el{I wholeheartedly agree with your analysis, but feel like it’s ignoring the elephant in the room:} writing code is not the bottleneck in need of optimization. Conceiving the solution is. Any time ``saved'' through Copilot and it’s ilk is immediately nullified by having to check it's correctness. \el{From there, the problem is worsened by the Frankensteinesque stitching together of disparate parts that you describe.
    I can’t imagine how Copilot would save anything but a negligible amount of effort for someone who is actually thinking about what they’re writing.}''
\end{quote}

\begin{quote}
    ``What I want is a copilot that finds errors \el{ala spellcheck-esque. Did I miss an early return? For example in the code below

    def some_worker
        if disabled_via_feature_flag
             logger.info("skipping some_worker")
        some_potentially_hazardous_method_call()

    Right after the logger call I missed a return. A copilot could easily catch this.} Invert the relationship. I don't need some boilerplate generator, I need a nitpicker that's smarter than a linter. I'm the smart thinker with a biological brain that is inattentive at times. Why is the computer trying to code and leaving mistake catching to me? It's backwards.''
\end{quote}

\begin{quote}
    ``I turned off auto-suggest and that made a huge difference. Now I'll use it when I know I'm doing something repetitive that it'll get easily, or if I'm not 100\% sure what I want to do and I'm curious what it suggests. This way I get the help without having it interrupt my thoughts with its suggestions.''
\end{quote}

Another frequent experience is that language models can introduce subtle, difficult to detect bugs, which are not the kind that would be introduced by a human programmer writing code manually. Thus, existing developer intuitions around the sources of errors in programs can be less useful, or even misleading, when checking the correctness of generated code. 

One developer reported their experience of having an incorrect, but plausible-sounding field name suggested by Copilot (\texttt{accessTokenSecret} instead of \texttt{accessSecret}) and the consequent wild goose chase of debugging before discovering the problem. As sources of error, these tools are new, and developers need to learn new craft practices for debugging. \emph{``There are zero places that can teach you those things. You must experience them and unlock that kind of knowledge.''}, the developer concludes, \emph{``Don't let code completion AI tools rule your work. \el{I'm sure there is a lot of funny stories like this. But we all make mistakes. Embrace them as learning opportunities.} I don't blame [Copilot] for this. I blame myself. But whatever. At least I got some experience.''}. Commenters on Hacker News report similar experiences:

\begin{quote}
    ``\el{I've been using Copilot non-stop on every hobby project I have ever since they've let me in (2021/07/13) and I am honestly flabbergasted they think it's worth 10\$/mo. My experience using it till this day is the following:
    - It's an amazing all-rounder autocomplete for most boilerplate code. Generally anything that someone who's spent 5 minutes reading the code can do, Copilot can do just as well.

    - It's terrible if you let it write too much.} The biggest problem I've had is not that it doesn't write correctly, it's that it think it knows how and then produce good looking code at a glance but with wrong logic. \el{

    - Relying on its outside-code knowledge is also generally a recipe for disaster: e.g. I'm building a Riichi Mahjong engine and while it knows all the terms and how to put a sentence together describing the rules, it absolutely doesn't actually understand how "Chii" melds work
    
    - Due to the licensing concerns I did not use CoPilot at all in work projects and I haven't felt like I was missing that much. A friend of mine also said he wouldn't be allowed to use it.
    
    You can treat it as a pair programming session where you're the observer and write an outline while the AI does all the bulk work (but be wary), but at what point does it become such a better experience to justify 10\$/mo? I don't understand if I've been using it wrong or what.}''
\end{quote}

\begin{quote}
    ``\el{The more experience I get with GPT-3 type technologies, the more I would never let them near my code. It wasn't an intent of the technology per se, but} it has proved to be very good at producing superficially appealing output that can stand up not only to a quick scan, but to a moderately deep reading, but still falls apart on a more careful reading. \el{At least when that's in my prose it isn't cheerfully and plausibly charging the wrong customer or cheerfully and plausibly dereferencing a null pointer.

    Or to put it another way,} it's an uncanny valley type effect. \el{All props and kudos to the technologists who developed it, it's a legitimate step forward in technology, but at the same time} it's almost the most dangerous possible iteration of it, where it's good enough to fool a human functioning at anything other than the highest level of attentiveness but not good enough to be correct all the time. See also, the dangers of almost self-driving cars; either be self-driving or don't but don't expect halfway in between to work well.''
\end{quote}

\begin{quote}
    ``\el{I turned off copilot a week ago for the same reason.} The code it generates \_looks\_ right but is usually wrong in really difficult to spot ways but things you’d never write yourself.''
\end{quote}

Many developers reported concerns around such tools repeating private information, or repeating copyrighted code verbatim, which might have implications for the licenses in their own projects. Notions of the dangers of such ``stochastic parrots'' \citep{DBLP:conf/fat/BenderGMS21} are not new and have been well-explored, and are not as directly connected to the user experience of programming assistance as some of the other concerns we have listed here. As such, we will not enter that discussion in depth here, except to mention that these concerns were present in several blog articles and online discussions.

Thus, in practice, programmers describe the challenges of writing effective prompts, misinterpreted intent, code that includes subtle bugs or poor programming practices, the burden of inspecting and checking that generated code is correct, and worries about private information, plagiarism and copyright.

\subsection{These tools are useful for boilerplate and code reuse}
Despite the challenges we have described so far in this section, the utility of these tools in certain contexts is undeniable, and some programmers report having developed workflows, in certain contexts, that are heavily dependent on AI assistance. Particularly for simple tasks that require a lot of ``boilerplate'' code, or common tasks for which there are likely to be snippets of code online which prior to these AI assistants would have required a web search to retrieve. Hacker News commenters write:

\begin{quote}
``These days not having Copilot is a pretty big productivity hit to me. The other day Copilot somehow stopped offering completions for maybe an hour, and I was pretty shocked to realize how much I've grown to rely on just hitting tab to complete the whole line. (I was writing Go at the time which is on the boilerplatey side among the mainstream languages, so Copilot is particularly effective \el{there. Auto-completion from gopls — the language server — is no match.) I'm officially spoiled.

I'm afraid small independent code editors are increasingly fighting uphill battles as the big ones roll out support for more and more productivity boosts like LSP and Copilot integration.}''
\end{quote}

\begin{quote}
    ``I use GTP-3 codex [sic] daily when working. It saves me time, helps me explore unfamiliar languages and APIs and generates approaches to solve problems. It can be shockingly good at coding in narrow contexts. It would be a mistake to miss the developments happening in this area''
\end{quote}

\begin{quote}
``\el{This is somewhat tangential but} for a lot of quick programming questions, I'm finding I don't even need a search engine.
I just use Github Copilot. For example, if I wanted to remember how to throw an exception I'd just write that as a comment and let Copilot fill in the syntax. Between that and official docs, don't need a ton else.''
\end{quote}

\begin{quote}
``\el{This is spot on, and I've found myself making the same argument recently about GitHub's Copilot.} It's changing the way I write code in a way that I can already tell is allowing me to be much lazier than I've previously been about learning various details of languages and libraries. \el{There's an obvious tendency to play Thamus and judge that as a bad thing, but I think the reality is that Copilot and future generations of tools inspired by it will completely change the operation of programming in ways we can't even imagine yet, allowing us to build software more quickly than we currently suspect.
It's a train of thought that spans the arc of history: the New Thing, which allows young people to get by without developing deep skills in an area that was previously critical to success, is inherently bad because it enable laziness. In reality, it provides the leverage necessary to do things faster, or sometimes to do previously-unimaginable things.}''
\end{quote}

\begin{quote}
``\el{Well, yes.
There's } Github Copilot \el{, which} pretty much replaced almost my entire usage of Stack Overflow.\el{

But the thing is that with a rando package is that one doesn't have to review the code. Sure, the code is also coming from somewhere else. Sure, it might never get updated. Sure it might be full of bugs. Sure, it might be more dangerous than copying from StackOverflow. But out of sight, out of mind.

In the end the overuse of packages isn't about saving time or "doing the best for business" or "ensuring that the code is maintained by someone else". It's purely about covering our asses}''
\end{quote}

\begin{quote}
``\el{I would disagree. Just like an airplane copilot,} GitHub Copilot really shines in rote work: when it can correctly infer what you are about to do, it can and will assist you correctly. It’s not able to make big decisions, but in a pinch, it might be able to give hints. \el{
In short: Write your class methods or functions headers-first and then start working on the bodies. Add step-by-step comments to a body before starting to add code. These are, of course, good habits to foster even without Copilot.

Copilot will pick up what you do in the first method/function/whatever structure and suggest adapted completions for the rest.

In general, when there are multiple similar things to write, name your cases, then write one of them in detail, move to the next empty one and watch magic happen.

} If used right, Copilot can give developers a significant velocity boost, especially in greenfield projects where there is lots and lots of boilerplate to write. \el{

It is still a ”copilot” though so keep an eye on it, lots of ”wtf is that guy doing” moments ahead for sure.}''
\end{quote}

%% file: sec_related_models.tex
\section{The inadequacy of existing metaphors for AI-assisted programming}
\label{sec:metaphors}

\subsection{AI assistance as search}

In research studies, as well as in reports of developer experiences, comparisons have been drawn between the nature of AI programming assistance and programming by searching and reusing code from the Internet (or from institutional repositories, or from the same project, or from a developer's previous projects). 

The comparison between AI programming assistance and search is a natural one, and there are many similarities. Superficially, both have a similar starting point: a \emph{prompt} or query that is predominantly natural language (but which may also contain code snippets). From the user perspective, both have an \emph{information asymmetry}: the user does not know precisely what form the result will take. With both search and AI assistance, for any given query, there will be \emph{several results}, and the user will need to invest time evaluating and comparing them. In both cases, the user may only get an \emph{inexact solution}, or indeed nothing like what they want, and the user may need to invest time adapting and repairing what they get.

However, there are differences. When searching the web, programmers encounter not just code, but a variety of types of results intermingled and enmeshed. These include code snippets interspersed with human commentary, perhaps discussions on forums such as Stack Overflow, videos, and images. A search may return new APIs or libraries related to the query, thus showing results at different levels of abstraction. Search has signals of provenance: it is often (though not always) possible to determine the source of a code snippet on the web. There is a lot of information scent priming to assist with the information foraging task \citep{srinivasa2016foraging}. In this way, programming with search is a \emph{mixed media} experience.


In contrast, programming with large language models can be said to be a \emph{fixed media} experience. The only output is tokens (code, comments, and data) that can be represented within the context of the code editor. This has some advantages: the increased speed of code insertion (which is the immediate aim) often came up in experience reports. However, the learning, exploration, and discovery, and access to a wide variety of sources and media types that occurs in web search is lost. Provenance, too is lost: it is difficult to determine whether the generation is original to the model, or a stochastic parroting \citep{DBLP:conf/fat/BenderGMS21, ziegler_2021}. Moreover, due to privacy, security, and intellectual property concerns, the provenance of code generated by large language models may be withheld or even destroyed \citep{sarkar2022explainable}. This suggests that in future assistance experiences, mixed-media search might be integrated into programmer assistance tools, or the models themselves might be made capable of generating more types of results than the simple code autocomplete paradigm of current tools.



\subsection{AI assistance as compilation}
An alternative perspective is that AI assistance is more like a compiler. In this view, programming through natural language prompts and queries is a form of higher-level specification, that is `compiled' via the model to the source code in the target language, which is lower level.

Let us (crudely) assume that as programming notations travel along the abstraction continuum from `lower' to `higher' levels, the programmer becomes, firstly, less concerned with the mechanistic details of program execution, and secondly, more and more declarative, specifying \emph{what} computation is required rather than \emph{how} to compute it. In general, these are desirable properties of programming notations, but they do not always make the activity of programming easier or more accessible. As people who write code in declarative languages or formal verification tools will tell you, it's often much more difficult to specify the \emph{what} than the \emph{how}. The much more broadly adopted practice of test-driven development is adjacent; while tests are not necessarily written in a higher-level language than the code, they aim to capture a higher-level notion of correctness, the \emph{what} of the problem being solved. Learning to be a test engineer takes time and experience, and the entire distinct career path of ``software engineer in test'' attests to the specialised requirements of programming at higher levels of abstraction.

Some would draw a distinction between programming in a specification language and a compiled programming language. Tony Hoare himself considers these different, on the grounds that while a compiler only aims to map a program from the source language into a finite set of valid programs in the target language, a specification might be satisfied by an infinite number of valid programs (\emph{pers comm.}, first author, ca. 2014). Thus the technical and interaction design problems of programming through specification refinement encompasses, but is much broader than, the technical and interaction design problems of compilers. While we acknowledge this distinction, there is insufficient empirical evidence from the experience reports summarised in Section~\ref{sec:experience_reports} that working programmers themselves consistently make a meaningful distinction between these concepts.



Programming with large language models, like in a higher-level notation, also allows the programmer to be less concerned with details of the target language. For example, developers in our experience reports relied on AI assistance to fill in the correct syntax, or to discover and correctly use the appropriate API call, thus allowing them to focus on higher-level aspects of the problem being solved. However, there are fundamental differences between this experience and the experience of using a compiler. First, the abstraction is not complete, i.e., a programmer cannot \emph{completely} be unaware of the target language, they must still be able to understand and evaluate the generated code in order to use such tools effectively. With compilers, although knowledge of the target language can help experienced developers in certain circumstances, it is far from a prerequisite for effective usage. Moreover, compilers can be relied on almost universally to generate a correct and complete translation from source to target language, whereas programming with AI assistance involves the active checking and adaptation of translated code. Next, compilers are (comparatively) deterministic, in that they consistently produce the same output for the same input, but this is not the case for current AI programming tools (although this is not a fundamental limitation, and consistency can be enforced). Finally, though they are often criticised for being cryptic and unhelpful \citep{barik2018should}, compilers do offer levels of interaction and feedback through warnings and error messages, which help the programmer improve the code in the source language; there is currently no such facility with AI programming tools and this strikes us as an area with potential for innovation.

Perhaps more profoundly, while natural language can be used to express concepts at a higher abstraction level, the \emph{range} of abstraction expressible in natural language is much wider than with other forms of programming notation. Traditional programming notations with ad-hoc abstraction capabilities (subroutines, classes, etc.) allow programmers to manually raise the level of abstraction of their own code and APIs. But with code generated by language models, as we have seen from the reports in Section~\ref{sec:experience_reports}, a prompt can span the gamut from describing an entire application in a few sentences, to painstakingly describing an algorithm in step-by-step pseudocode. Thus it would be a mistake to view programming with AI assistance as another rung on the abstraction ladder. Rather, it can be viewed as a device that can teleport the programmer to arbitrary rungs of the ladder as desired.

We close the discussion on AI assistance as a compiler with a few miscellaneous notes. The idea of using natural language as a programming notation has a long history (e.g., \citep{miller1981natural, lieberman2006feasibility}), which we will not cover here. However, it is notable that there are many ways that natural language has been integrated with programming, such as debugging \citep{ko2004designing}. With large language models, there are better capabilities for inference of intent and translation to code, but therefore also the potential to open up new strategies for inspecting and explaining code. There are also new failure modes for this paradigm of programming.


\subsection{AI assistance as pair programming}
The third common perspective is that AI-assisted programming is like pair programming. GitHub Copilot's commercial tagline describes it as ``your AI pair programmer''. As opposed to search and compilation, which are both relatively impersonal tools, the analogy with pair programming is evocative of a more bespoke experience; assistance from a partner that understands more about your specific context and what you're trying to achieve. AI-assisted programming does have the potential to be more personalised, to the extent that it can take into consideration your specific source code and project files. As Hacker News commenters write:

\begin{quote}
    ``\el{I've had this experience too. Usually it's meh, but} at one point it wrote an ENTIRE function by itself and it was correct. \el{IT WAS CORRECT! And} it wasn't some dumb boilerplate initialization either, it was actual logic with some loops. The context awareness with it is off the charts sometimes.\el{

    Regardless I find that while it's good for the generic python stuff I do in my free time, for the stuff I'm actually paid for? Basically useless since it's too niche. So not exactly worth the investment for me.}''
\end{quote}

\begin{quote}
    ``\el{I've been using Copilot for a few months and...
    Yeah, it makes mistakes, sometimes it shows you i.e. the most common way to do something, even if that way has a bug in it.
    
    Yes, sometimes it writes a complete blunder.
    
    And yes again, sometimes there are very subtle logical mistakes in the code it proposes.
    
    But overall? It's been *great*! Definitely worth the 10 bucks a month (especially with a developer salary). :insert shut up and take my money gif:
    
    It's excellent for quickly writing slightly repetitive test cases; it's great as an autocomplete on steroids that completes entire lines + fills in all arguments, instead of just a single identifier; it's great for quickly writing nice contextual error messages (especially useful for Go developers and the constant errors.Wrap, Copilot is really good at writing meaningful error messages there); and it's also great for technical documentation, as it's able to autocomplete markdown (and it does it surprisingly well).} It's like having the stereotypical ``intern'' as an associate built-in to your editor. \el{
    
    And sometimes, fairly rarely, but it happens, it's just surprising how good of a suggestion it can make.
    
    } It's also ridiculously flexible. When I start writing graphs in ASCII (cause I'm just quickly writing something down in a scratch file) it'll actually understand what I'm doing and start autocompleting textual nodes in that ASCII graph.''
\end{quote}

Besides personalisation, the analogy also recalls the conventional role-division of pair programming between ``driver'' and ``navigator''. When programming, one needs to form mental models of the program at many layers: from the specific statement being worked on, to its context in a subroutine, to the role that subroutine plays in a module, to the module within the program. However, code must be written at the statement level, which forces developers to keep this lowest level constantly at the forefront of their working memory. Experienced developers spend more time mapping out their code so that they can spend less time writing it. Research into code display and navigation has explored how different ways of presenting lines of code can help programmers better keep these different layers of mental models in mind \citep{henley2014patchworks}. Pair programming, the argument goes, allows two partners to share the burden of the mental model. The driver codes at the statement and subroutine level while the navigator maps out the approach at the module and program level. 

By analogy to pair programming, the AI assistant taking the role of the driver, a solo programmer can now take the place of the navigator. But as we have seen, the experience of programming with AI assistance does not consistently absolve the human programmer of the responsibility for understanding the code at the statement and subroutine level. The programmer may be able to become \emph{``lazier [...] about learning various details of syntax and libraries''}, but the experience still involves much greater statement-level checking. 

While a pair programming session requires a conscious, negotiated decision to swap roles, a solo programmer with an AI assistant might find themselves fluidly traversing the spectrum from driving to navigation, from one moment to the next. This may partially explain why, in a preliminary experiment (n=21) comparing the experience of ``pair programming'' with GitHub Copilot to programming in a human pair either as driver or navigator, \cite{imai2022github} finds that programmers write more lines of code with Copilot than in a human pair, but these lines are of lower quality (more are subsequently deleted).

Moreover, meta-analyses of pair programming have shown mixed efficacy of human pair programming on task time, code quality and correctness \citep{salge2016pair,hannay2009effectiveness}, suggesting that emulating the pair programming experience is not necessarily a good target to aim for. Multiple studies have concluded that the apparent successes of pair programming can be attributed, not to the role division into driver and navigator, but rather the high degree of \emph{verbalisation} that occurs when pair programmers are forced to rationalise their decisions to each other \citep{hannay2009effectiveness}. Others have found that programming in pairs induces greater focus out of a respect for shared time; pair programmers are less likely to read emails, surf the web, or take long phone calls \citep{williams2000all}. These particular benefits of pair programming are not captured at all by AI assistance tools.


The comparison to pair programming is thus relatively superficial, and today's experience of AI-assisted programming is not comparable with pair programming to the same extent as it is with search or compilation.

\subsection{A distinct way of programming}

LLM-assisted programming assistance bears similarities to search: both begin with a prompt, both have an information asymmetry, there are several results, with inexact solutions. But there are differences: search is mixed-media, whereas LLM assistance is fixed. Search (often) has provenance, and language models do not.

It also bears similarities to compilation and programming by specification. Both enable programming at a `higher' level of abstraction (for some definition of higher). Yet unlike with compilers, a programmer using AI assistance must still have a working knowledge of the target language, they must actively check the output for correctness, and they get very little feedback for improving their `source' code.

It also bears a superficial similarity to pair programming, in that it promises to let the programmer take the role of `navigator', forming high-level mental models of the program while delegating the role of `driver' to the language model. But unlike with pair programming, the human navigator must often hop into the driver's seat. And unlike with pair programming, LLM-assisted programming does not require verbalisation, nor does it coerce greater focus out of a respect for shared time.

Thus existing metaphors do not completely capture the experience of LLM-assisted programming. It is emerging as a distinct way of programming. It does not quite strike us as a distinct \emph{practice} of programming, as that term has been applied to communities of programmers united by similar ethos and aims, such as enterprise software engineers, bricoleurs, live coders, and code benders; but as \cite{bergstrom2016practices} note, there are no clear criteria by which we can define the boundaries of a practice. Nor does it strike us as being a new \emph{activity} of programming as per the cognitive dimensions framework, since AI assistance is clearly orthogonal to authoring, transcription, and modification, being applicable to each of these activities and others besides. Yet as a way of programming it seems to affect programmer's experience more profoundly than a feature such as autocomplete, having far-reaching impact on their attitudes and practices of authoring, information foraging, debugging, refactoring, testing, documentation, code maintenance, learning, and more.




%% file: experience_report_sources.tex
\appendix
\section{Experience report sources}
\label{apx:experience_report_sources}
This appendix contains a list of sources we draw upon for the quotes and analysis in Section~\ref{sec:experience_reports}. While all sources were included in our analysis, we did not draw direct quotes from every source in this list.

\subsection{Blog posts and corresponding Hacker News discussions}
\begin{enumerate}
    \item Andrew Mayne, March 17 2022, ``Building games and apps entirely through natural language using OpenAI’s code-davinci model''. 
            URL: \url{https://andrewmayneblog.wordpress.com/2022/03/17/building-games-and-apps-entirely-through-natural-language-using-openais-davinci-code-model/}. 
            Hacker News discussion: \url{https://news.ycombinator.com/item?id=30717773}
    \item Andrew Mouboussin, March 24 2022, ``Building a No-Code Machine Learning Model by Chatting with GitHub Copilot''. 
            URL: \url{https://www.surgehq.ai/blog/building-a-no-code-toxicity-classifier-by-talking-to-copilot}. 
            Hacker News discussion: \url{https://news.ycombinator.com/item?id=30797381}
    \item Matt Rickard, August 17 2021, ``One Month of Using GitHub Copilot''. 
            URL: \url{https://matt-rickard.com/github-copilot-a-month-in/}.
    \item Nutanc, November 15 2021, ``Using Github copilot to get the tweets for a keyword and find the sentiment of each tweet in 2 mins''. 
            URL: \url{https://nutanc.medium.com/using-github-copilot-to-get-the-tweets-for-a-keyword-and-find-the-sentiment-of-each-tweet-in-2-mins-9a531abedc84}.
    \item Tanishq Abraham, July 14 2021, ``Coding with GitHub Copilot''. 
            URL: \url{https://tmabraham.github.io/blog/github_copilot}.
    \item Aleksej Komnenovic, January 17 2022, ``Don't fully trust AI in dev work! /yet''. URL: \url{https://akom.me/dont-fully-trust-ai-in-dev-work-yet}.
\end{enumerate}

\subsection{Miscellaneous Hacker News discussions}
\begin{enumerate}
    \item\url{https://news.ycombinator.com/item?id=30747211}
    \item\url{https://news.ycombinator.com/item?id=31390371}
    \item\url{https://news.ycombinator.com/item?id=31020229&p=2}
    \item\url{https://news.ycombinator.com/item?id=29760171}
    \item\url{https://news.ycombinator.com/item?id=31325154}
    \item\url{https://news.ycombinator.com/item?id=31734110}
    \item\url{https://news.ycombinator.com/item?id=31652939}
    \item\url{https://news.ycombinator.com/item?id=30682841}
    \item\url{https://news.ycombinator.com/item?id=31515938}
    \item\url{https://news.ycombinator.com/item?id=31825742}
\end{enumerate}